\documentclass[aps,12pt]{revtex4}
 \usepackage{graphicx}
\usepackage{color}
\usepackage{float}

\begin{document}

\title{A new type of half-quantum circulation in a macroscopic polariton spinor ring condensate}

\author{Gangqiang Liu$^1$,
David W. Snoke,$^1$\footnote{email: snoke@pitt.edu} 
Andrew Daley$^2$, 
Loren Pfeiffer$^3$, and Kenneth West$^3$}

\affiliation{$^1$Department of Physics and Astronomy, University of Pittsburgh, Pittsburgh, PA 15260, USA\\
$^2$Department of Physics and SUPA, University of Strathclyde, Glasgow G4 0NG, Scotland, UK\\
$^3$Department of Electrical Engineering, Princeton University, Princeton, NJ 08544, USA}

\begin{abstract}
{We report the observation of coherent circulation in a macroscopic Bose-Einstein condensate of polaritons in a ring geometry. Because it is a spinor condensate, ``half quanta'' are allowed in which there is a phase rotation of $\pi$ in connection with a polarization vector rotation of $\pi$ around a closed path. This is clearly seen in the experimental observations of the polarization rotation around the ring. In our ring geometry, the half quantum state which we see is one in which the handedness of the spin flips from one side of the ring to the other, in addition to the rotation of the linear polarization component; such a state is allowed in a ring geometry but will not occur in a simply-connected geometry.  This state is lower in energy than a half quantum state with no change of the spin direction, and corresponds to a superposition of two different elementary half-quantum states.  The direction of circulation of the flow around the ring fluctuates randomly between clockwise and counterclockwise from one shot to the next; this corresponds to spontaneous breaking of time-reversal symmetry in the system. This new type of macroscopic polariton ring condensate allows for the possibility of direct control of the circulation to excite higher quantized states and creation of tunneling barriers to make the system resemble a SQUID.}
\end{abstract}

\maketitle

Ring condensates, analogous to superconducting rings, have received much attention lately \cite{ring1,ring1b,ring1c,ring2,ring3,ring4,cat-ring1,cat-ring1a,cat-ring2}; among other predictions, a ring condensate allows the possibility of macroscopic superposition of states with different circulation. A ring condensate is topologically distinct from a condensate in a simply connected region.  

With the advance of the field of polariton condensates in the past few years, it is a natural step to create a condensate ring in a microcavity polariton system. The polariton system allows direct, nondestructive observation of the momentum distribution, energy distribution, and spatial distribution of the particles, as well as direct measurement of the coherence properties via interferometry. To make a macroscopic ring requires macroscopic transport distances as well as macroscopic coherence length. This has been achieved with polaritons, with coherent motion over tens of microns and lifetimes of 10-20 ps \cite{bloch-wire,baum-prl2013}, and with coherent motion of polaritons over hundreds of microns with lifetimes of 150-200 ps \cite{prx,prb-mark,optica}. One advantage of the long-lifetime polariton systems is that the polaritons can move well away from the laser spot where they are generated, so that the laser can be viewed as a simple source term, and does not interact with the condensate.  For general reviews of previous polariton work with shorter transport distances, see Refs.~\cite{kav-book,german-book,deng,carusotto,pt}.

The polaritons can be viewed as photons which have been given a small effective mass, of the order of $10^{-4}$ times the mass of a vacuum electron, and repulsive interactions, which are about $10^4$ times stronger than the typical $\chi^{(3)}$ nonlinearities of photons in solids. The effective mass comes from the dispersion of the photons in a planar cavity, $\hbar\omega = \hbar c(k_{\|}^2+k_{\perp} )^{1/2}$, where $k_{\perp}$ is fixed by the width of the cavity, which implies $ \hbar\omega \simeq E_0 + \hbar^2k^2_{\|}/2m_{\rm eff}$, with $m_{\rm eff} = \hbar k_{\perp}/c$,  for low $k_{\|}$. There are two circular polarization modes of the cavity photons, corresponding to $m = \pm 1$ for the projection of the angular momentum on the $z$-axis perpendicular to the plane.  The strong interactions between photons are generated by mixing the photon states with a sharp excitonic resonance in a semiconductor inside the cavity, so that the photons pick up a fraction of the exciton-exciton interaction.  Although their interactions are much stronger than the interactions of typical photons in a solid medium, the polaritons are still in the weakly interacting Bose gas regime.

The structure for these experiments is a planar cavity in which the mirrors are distributed Bragg reflectors of AlAs/AlGaAs, and the exciton medium consists of GaAs/AlGaAs quantum wells embedded in this cavity. This is the same structure as that used in previous experiments, which allow coherent transport of polaritons over hundreds of microns in the two-dimensional plane of the cavity \cite{prx,prb-mark,optica}. Recent measurements \cite{optica} give the cavity lifetime as 135 ps, which corresponds to a polariton lifetime of 200 ps or more. While this may seem to be a short lifetime compared to atoms evaporating from an optical trap on time scales of seconds, the polariton lifetime is sufficient for them to interact many times with each other. In these new long-lifetime polariton systems, the ratio of lifetime in the trap to the particle-particle collision time can be of the order of 500:1, comparable to the ratio for cold atom condensates.

The lifetime of the polaritons and the strength of the interaction between the polaritons can be tuned by varying the energy difference between the photon states and the exciton states (known as the ``detuning''), which leads to a varying degree of mixing of the photons and excitons. Because the planar cavity has a wedge which gives a gradient of cavity width, we can tune the strength of the polariton-polariton interactions simply by choosing different locations on the sample with different cavity width.  There is a tradeoff in how much excitonic interaction character to give to the polaritons. Less interactions (more photon-like) allows long transport length, while more interactions allows better thermalization of the polariton gas via collisions and longer population lifetime.

\section{Creating the Ring Condensate}
For these experiments we chose a region of the sample in which the polaritons were slightly more photon-like than exciton-like. We then created an in-plane harmonic potential, using the inhomogeneous stress method that have previously been demonstrated \cite{balili-apl}. Because a shift of the exciton energy also affects the exciton-photon detuning and therefore the strength of the interaction between the polaritons, we must compensate the effect of the stress on the detuning by our initial choice of the detuning at the location in the cavity. In the experiment reported here, the detuning at the location of interest was -6.7 meV without stress, and +1.9 meV with the stress applied. 

We then created a Gaussian potential energy peak inside this harmonic potential using a laser focused to a spot, which generates an exciton cloud. The laser was non-resonant, with photon energy 105 meV higher than the polariton energy. This produced both excitons and polaritons at the laser focus spot. The excitons have a mass that is four orders of magnitude larger than the polaritons, and therefore they diffuse at most about 10 $\mu$m from their point of creation; they therefore act as a quasi-static barrier for the polaritons \cite{bloch-wire,baum-prl2013,prx}.  The sum of the harmonic potential and the Gaussian peak due to the exciton cloud makes a Mexican-hat potential. Figure~
1(a) shows our estimates for the different terms that contribute to the potential. Figure~
1(b) shows the intrinsic experimental energy profile recorded using a defocused laser, and Figure~
1(c) shows flow of the polaritons in the trap when the laser spot is focused, but not at the center of the trap, but instead is on the right side. The polaritons clearly flow away from the laser spot, about 35~$\mu$m, to the minimum of the harmonic potential. 

\begin{figure}
\begin{center}
\includegraphics[width=0.35\textwidth]{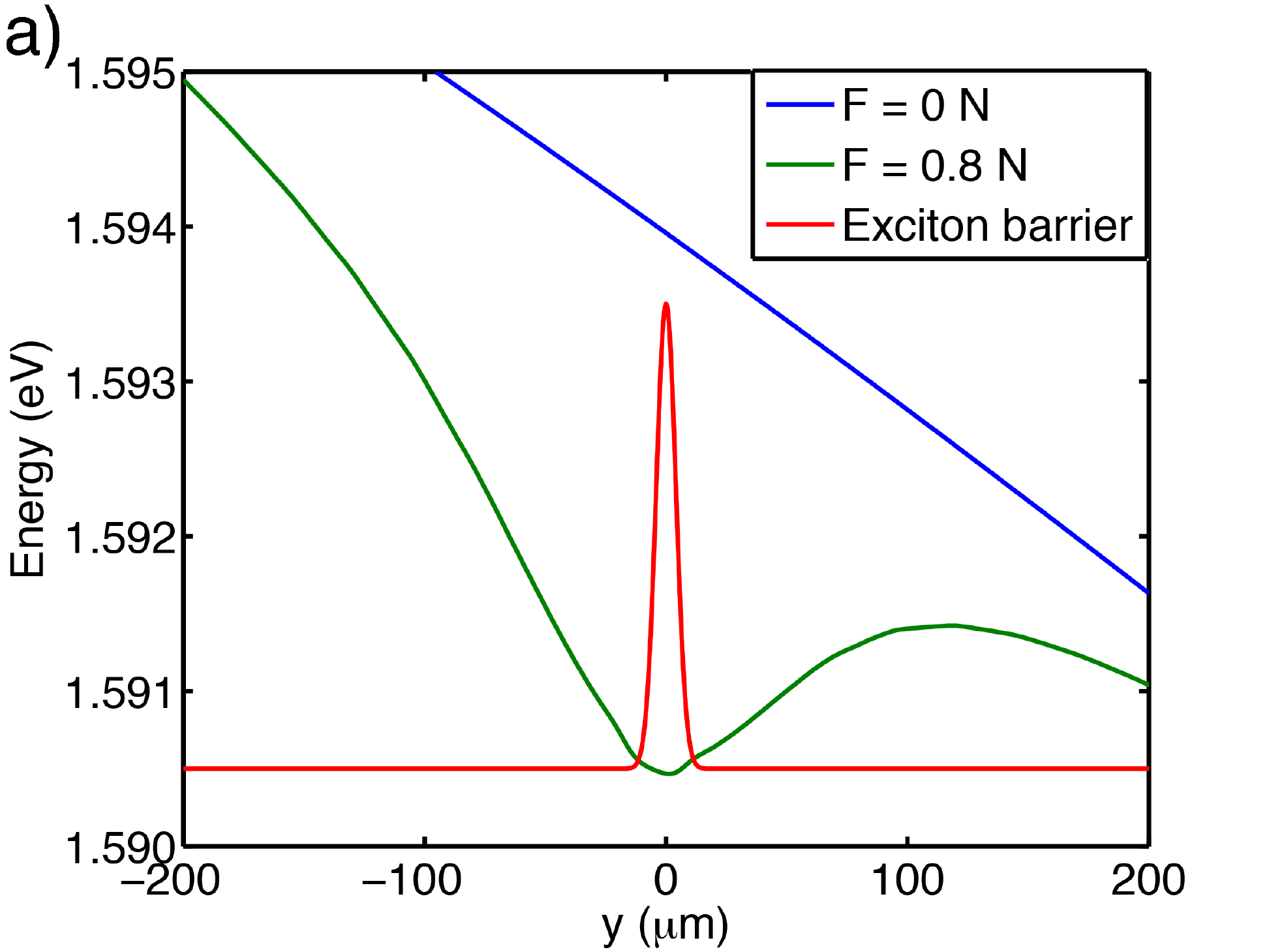}

\includegraphics[width=0.36\textwidth]{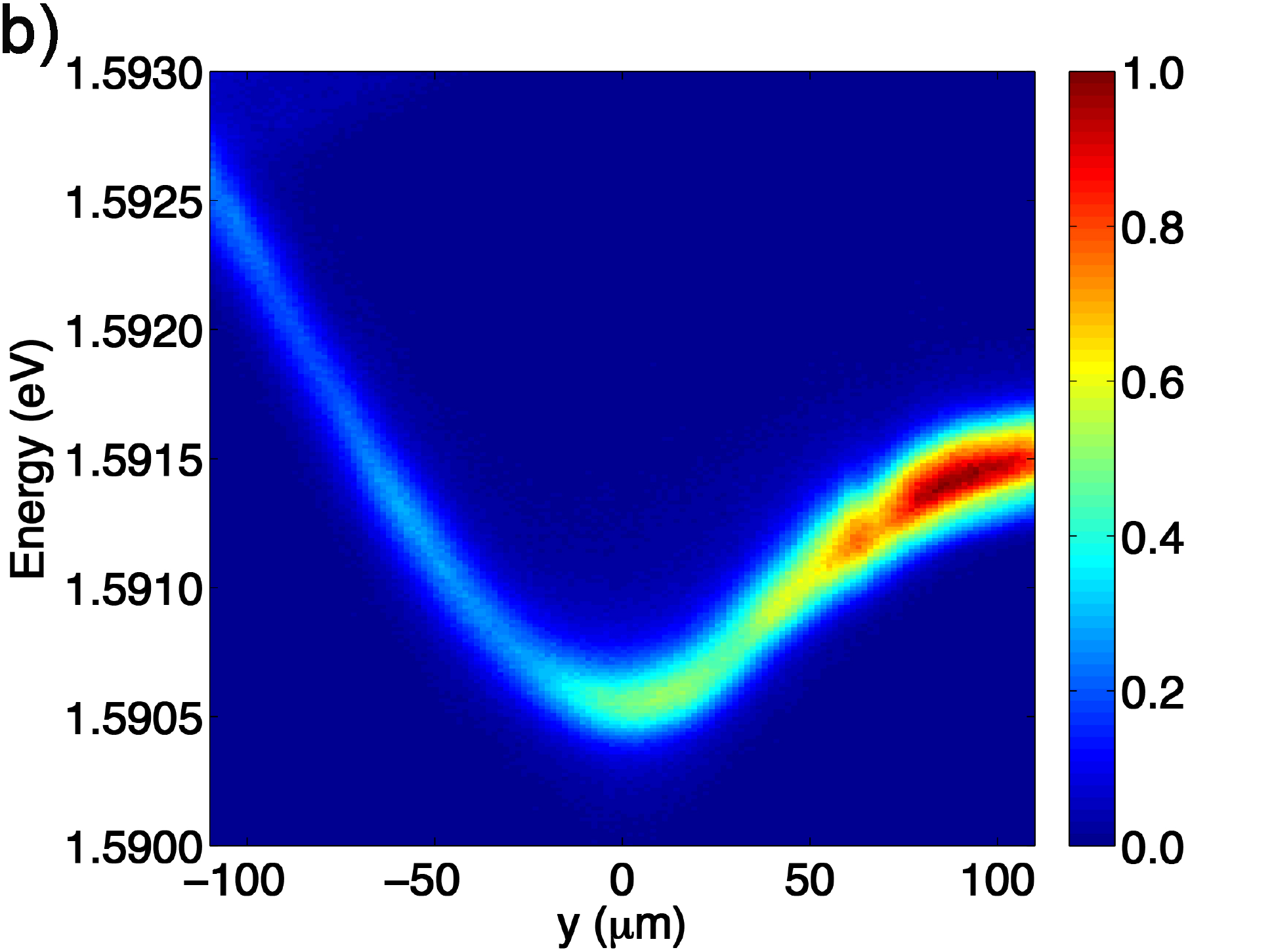}

\includegraphics[width=0.36\textwidth]{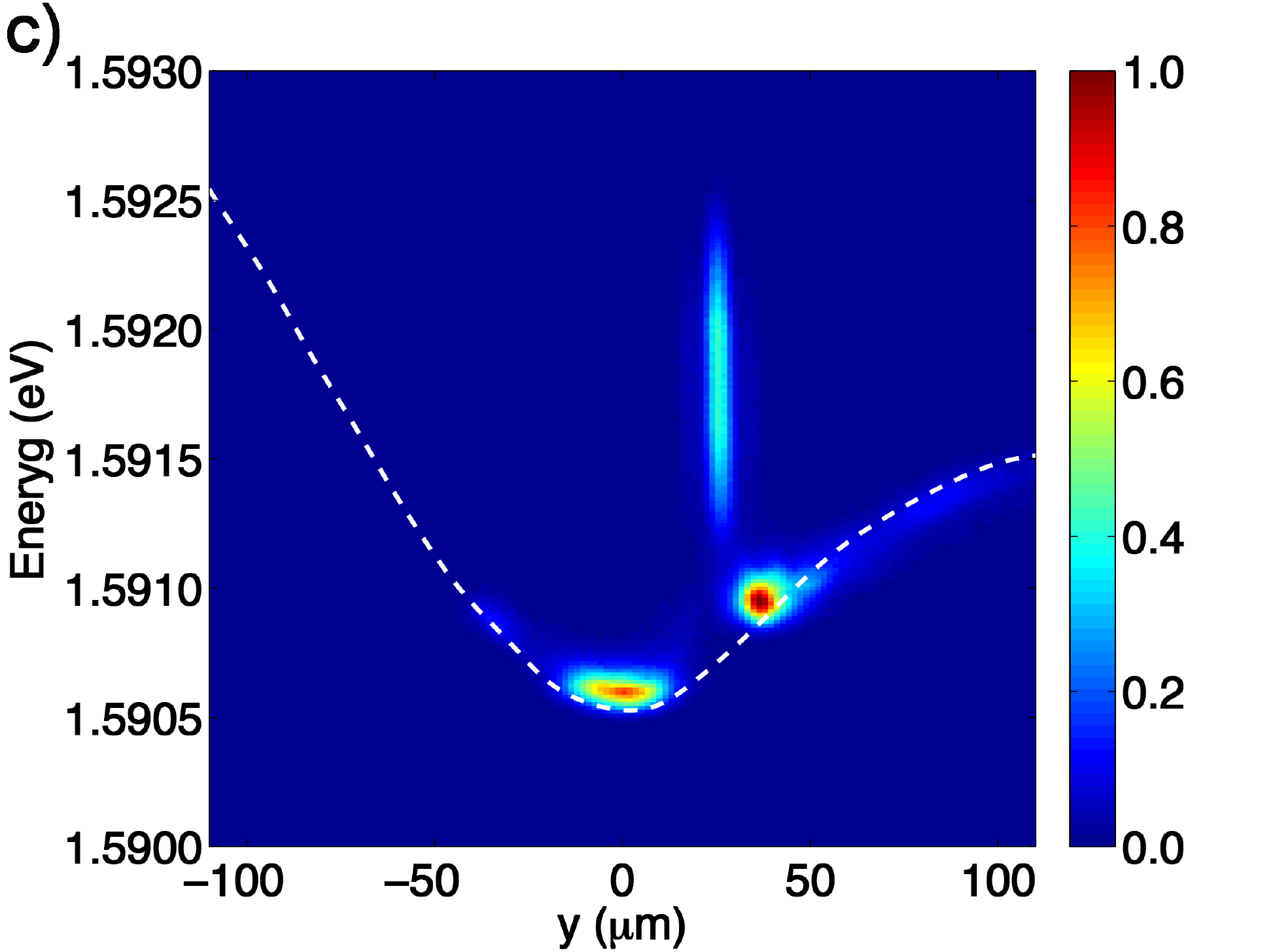}
\end{center}
\caption{ a) Plot of the three terms which contribute to the potential energy felt by the polaritons when they are condensed. Blue line: energy gradient due to the wedge in the cavity width (applied force on the stressor $F = 0$). Green line: the energy shifted by applied stress to have a harmonic potential minimum (applied force on the stressor $F=0.8$~N. Red line: potential peak created by a nearly-static cloud of excitons. b) The harmonic trap and gradient potentials, seen in the photon emission spectrum from the polaritons at very low density, generated using a defocused laser. (All color bars are normalized to maximum intensity $= 1.0$). c) The same system when the laser spot is tightly focused and set to one side instead of at the center of the trap. The vertical line is emission from polaritons at the laser spot, shifted up in energy by repulsion from the static exciton cloud. Polaritons flow to both sides of this spot, including to the global minimum.}
\label{potential}
\end{figure}

\begin{figure}
\begin{center}

\includegraphics[width=0.35\textwidth]{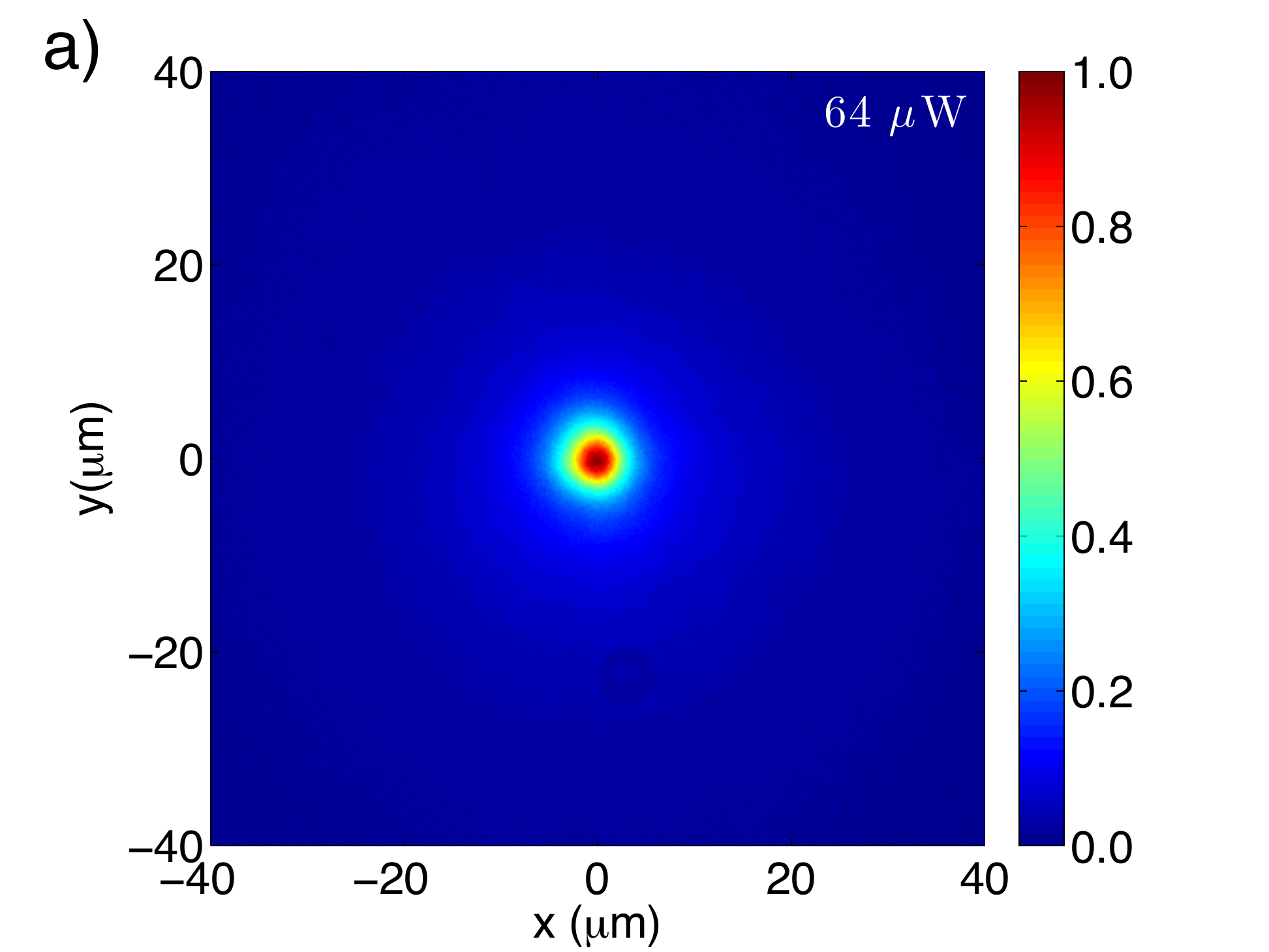}

\includegraphics[width=0.35\textwidth]{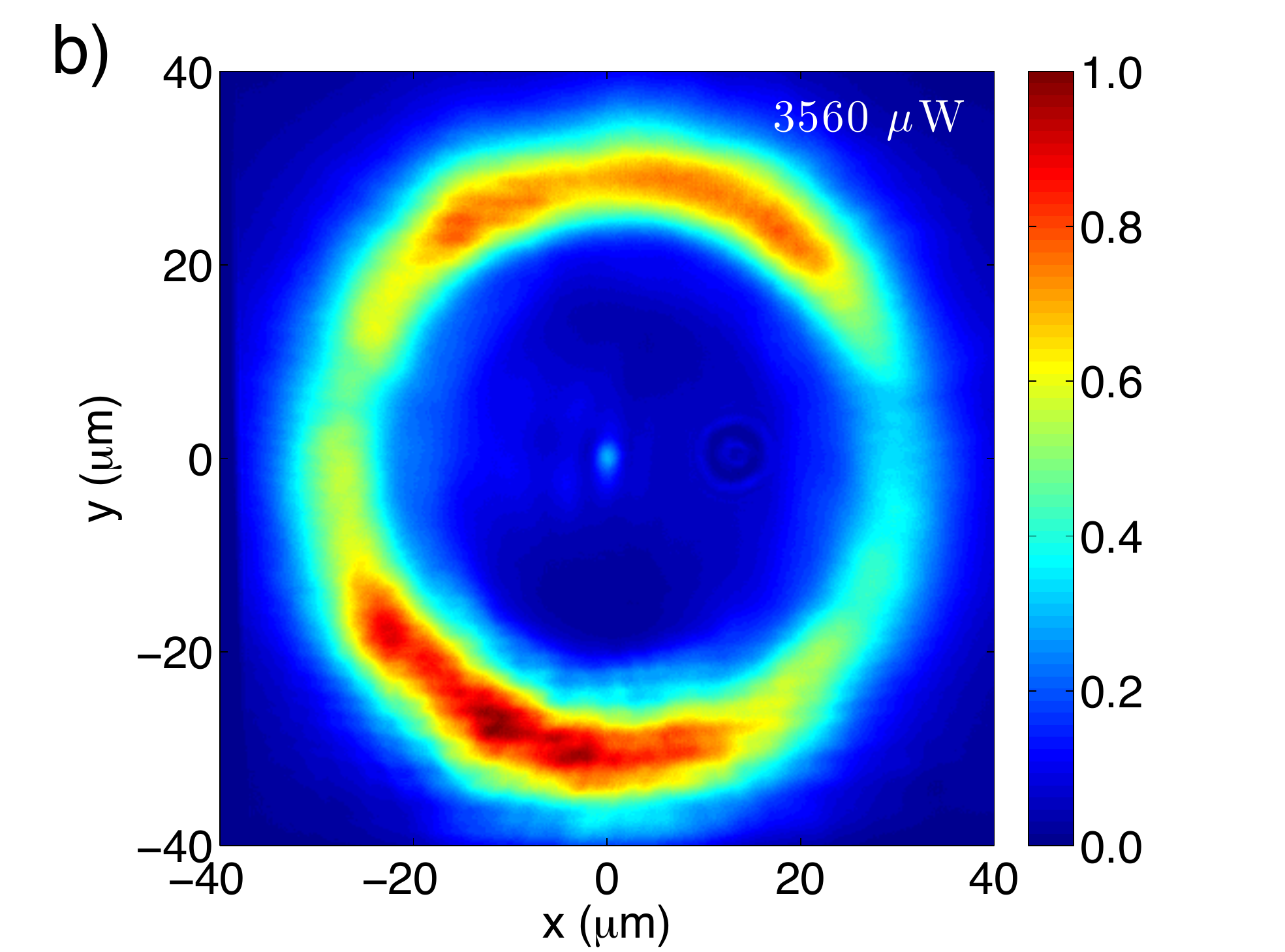}

\end{center}

\caption{a) Real-space image below critical threshold (64 $\mu$W average laser power, for a 1\% duty cycle with 25~$\mu$s duration pulses. b) Real-space image for the same conditions as (a), but above critical threshold for condensation (3.56 mW average laser power). The light polarizations are summed in these images.}
\label{ring}
\end{figure}

\begin{figure}
\begin{center}
\includegraphics[width=0.34\textwidth]{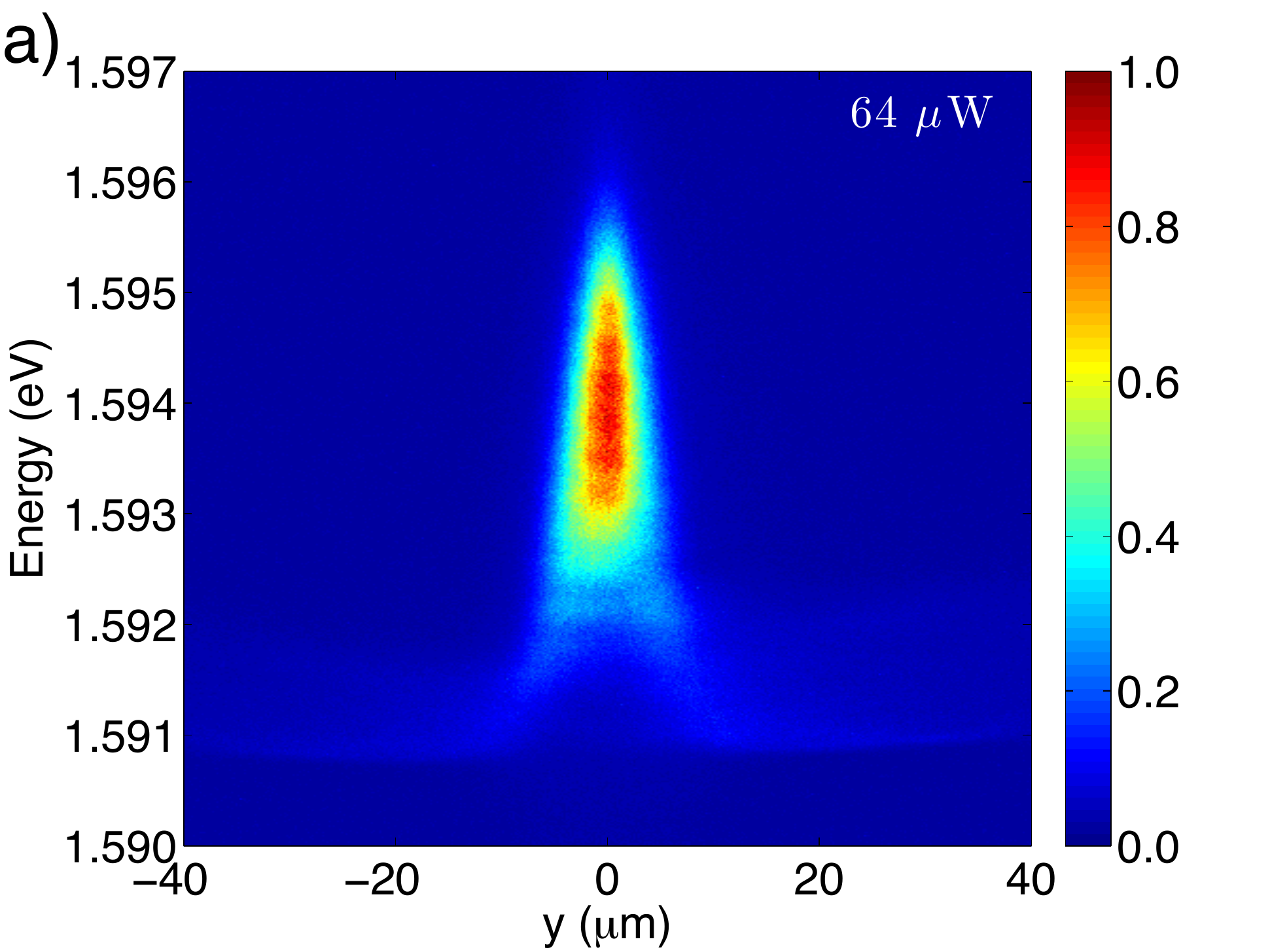}

\includegraphics[width=0.35\textwidth]{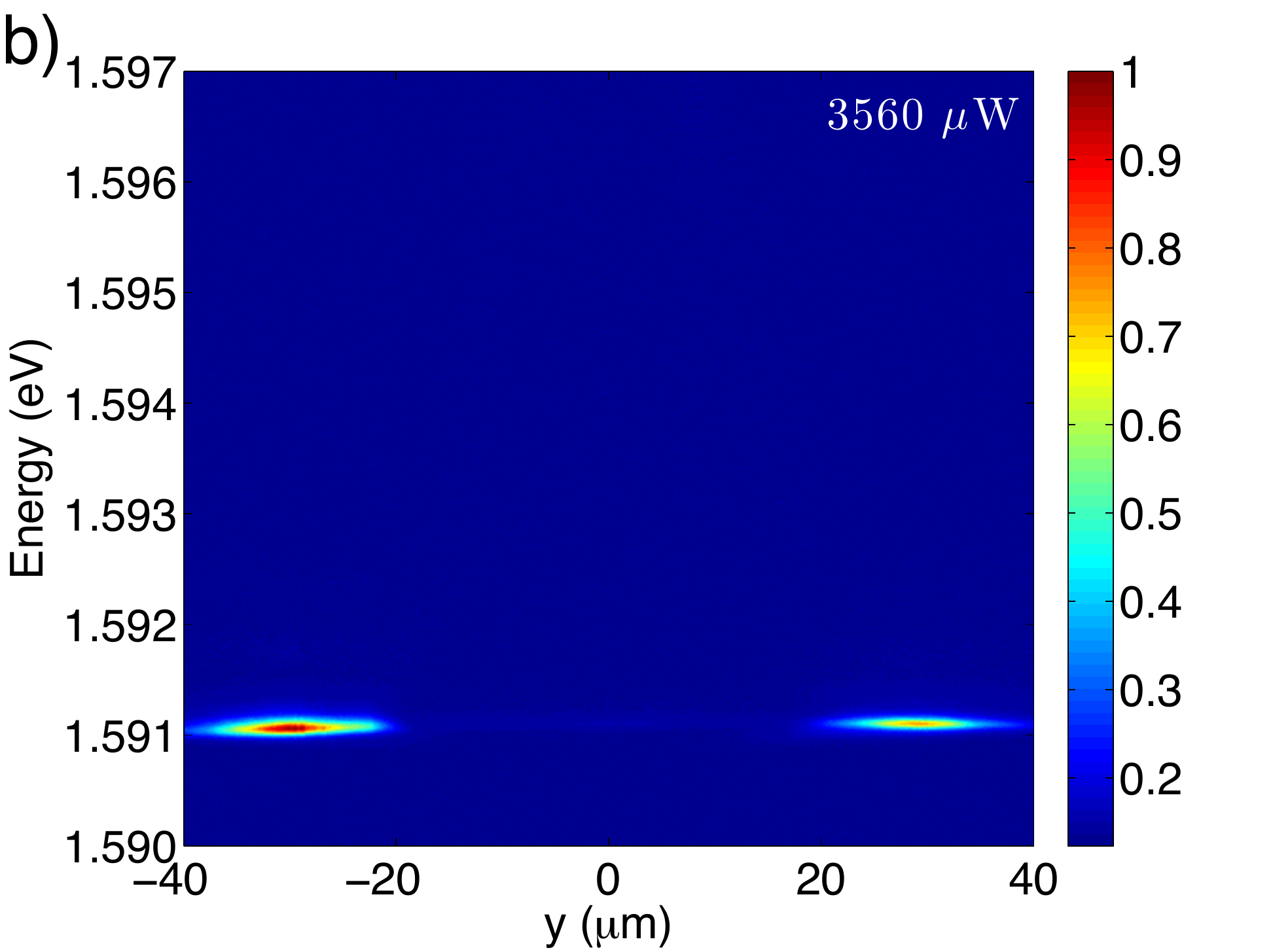}
\end{center}

 \caption{Spectral images along $y$, for $x=0$, for the same data as shown in Figure~2. a) Spectral image for $x=0$ for the conditions of Fig.~2(a). 
b)  Spectral image for $x=0$ for the same conditions as Fig.~2(b). }
\label{ring}
\end{figure}

To create the ring trap, we moved the laser focus to near the center of the harmonic potential minimum, as shown in Figure~
1(a). As in previous work \cite{prx}, when the density of the central exciton peak exceeds a critical value, there is a sharp transition to occupation of the ground state of the trap.  Figure~2 
shows the photon emission from the polaritons for two cases, below this critical density threshold and above the threshold. Figure~2(a) 
shows the emission from the laser focus when the density is below the critical density, and Figure~2(b) 
shows the filling of the ring above the critical density threshold. As seen in Figure~3(b), which gives the energy spectrum for the same conditions as Figure~2(b),  
the condensate is monoenergetic, with a narrow line width limited by our spectral resolution of 0.08~meV. Although there are density variations around the ring, the condensate fills in low-potential areas to maintain a single energy.  This ring condensate was observed under quasi-steady state conditions, with a continuous flow of polaritons generated at the central laser spot and flowing into the ring, which replace the loss of polaritons turning into external photons outside of the cavity, while the laser is on for a duration of 25~$\mu$s. A period of 2.5~ms with the laser off between pulses was used to prevent heating of the sample.

\begin{figure*}
\begin{center}
\includegraphics[width=0.95\textwidth]{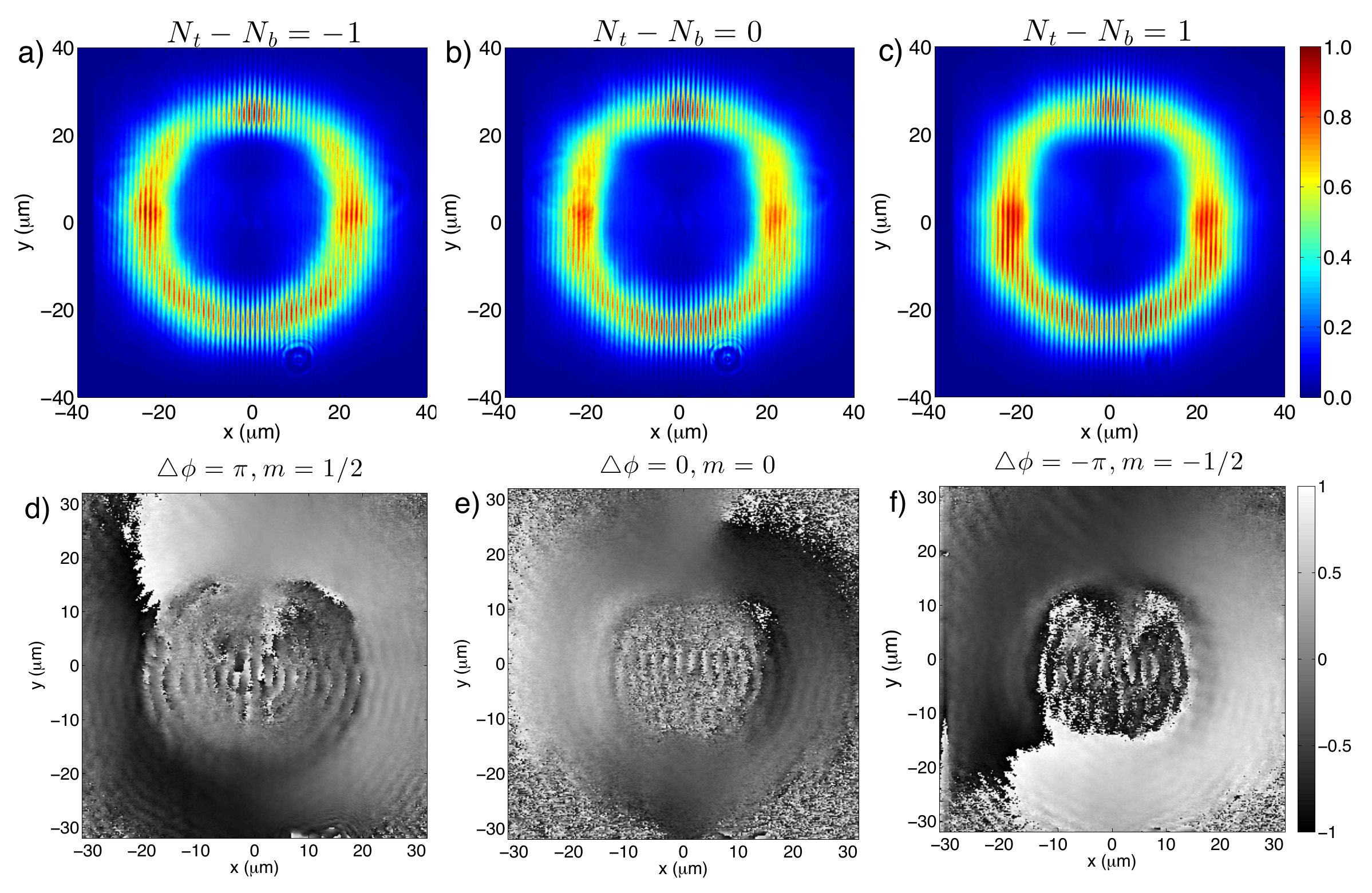}
\end{center}
 
\caption{a) Interference pattern between two images of the condensate, with one flipped across the $y$-axis, for average excitation laser power of 3.56 mW, in single shots with 25~$\mu$s duration. There is one more fringe on the bottom ($N_b$) than on the top ($N_t$) of the ring. b) Pattern for another single shot under the same conditions, with an equal number of fringes on the top and bottom of the ring. c) Pattern for another single shot with one more fringe on the top.  d) The phase map extracted from the interference pattern shown in (a). The circulation is counterclockwise. e) Phase map for the interference pattern of (b), showing no net circulation. f) Phase map for (c), showing clockwise circulation. The $\Delta \phi$ labels above the bottom row indicate the net phase shift around the ring. Since the interference pattern measures the phase shift of the ring condensate relative to itself in the opposite direction, the net phase shift around the ring is half of that shown in the phase map.}
\label{int}
\end{figure*}

The degree of spatial coherence of the condensate can be seen in an interference measurement. The top row of Figure~\ref{int} shows typical interference patterns when two copies of the spatial image of the condensate are overlapped, with one of the images flipped $x\rightarrow -x$. Fringes are seen across the entire image from one side to the other, showing that the coherence extends across the whole ring. 

A close analysis shows that the number of fringes on the top of the interference patterns in Fig.~\ref{int} is not always equal to the number of fringes on the bottom. In other words, there is net circulation in the condensate, defined as $\Gamma = \oint \vec{v}\cdot d\vec{l} $, which implies that the phase cannot be continuous---in the ring geometry, the potential barrier at the center of the ring makes the density of the condensate zero where the discontinuity occurs. The circulation can be seen in the phase maps which are produced from the interferometric data, shown as the lower row in Fig.~\ref{int}. 

The interference patterns are stable over the duration of the 25~$\mu$s pulses used here.  The pattern fluctuates from one pulse to the next, split with equal probability between a pattern with clockwise or counterclockwise circulation about 90\% of the time, and about 10\% of the time showing an equal number of fringes on the top and bottoms. We never see a difference of more than one fringe.

For the interference geometry we use here, a phase change of $2\pi$ around the ring in the phase map corresponds to a total phase change of only $\pi$ around the ring, since the interference pattern gives the phase change of the condensate relative to itself in opposite directions (see the supplemental material for discussion of this). These phase maps therefore indicate that the spinor nature of the polaritons, with two degenerate states, is important, since a scalar condensate must have a phase change of $2\pi n$ around a closed path.  

\section{Determining the Circulation State}

Figure 5(a) 
shows the direction of linear polarization at various points on the ring, deduced from measurements of the full Stokes vector of the light emitted at different points around the ring (see the supplementary material for the raw images of the Stokes-vector data). As seen in this figure, the linear polarization angle rotates by 180$^\circ$, while the circular component flips handedness on opposite sides of the ring. This is striking given that the underlying exciton states in GaAs-based structures have a fourfold symmetry, which is seen in a fourfold rotational symmetry of the polarization pattern under incoherent conditions \cite{nick-dark,pol-split}, and the eigenstates of the polariton states are linearly polarized\cite{pol-split}.  The orientation of the pattern is not connected to the underlying crystal symmetry; instead, it is fixed relative to the gradient of potential which exists in the system, which comes from the wedge in the cavity width.  The polarization pattern of the condensate also does not depend on the polarization of the laser which generates the polaritons at the central spot. 

The interference pattern and the polarization measurements can both be understood as the effects of quantized angular momentum in a spinor condensate. The generic effect of ``half quantization" has been worked out for spinor atom condensates \cite{atom-half}, $d$-wave superconductors \cite{SC-half} and in particular by Rubo \cite{rubo} for the case of polaritons in a simply-connected geometry; half-quantized vortices were reported experimentally for a polariton condensate localized in a sub-micron disorder minimum \cite{dev-half,dev-half2}. However, the state we see here is distinct from the half-vortex state of Rubo, and is favored only in a ring geometry. 

The Rubo state, when generalized to a ring geometry, consists of a $\pi$ rotation in phase around a closed path accompanied by a $\pi$ rotation of the polarization angle around the same path.  In terms of the linear polarization components in the plane of the sample, the azimuthal angle dependence of the Rubo state around a circle of constant radius can be written as \cite{rubo}
\begin{eqnarray}
\vec\varphi_{k,m} (\theta)  &= & \sqrt{n(\theta)}e^{im\theta}\left [f \left(
\begin{array}{cc}\cos(k\theta) \\
 \sin(k\theta) \end{array}
  \right) \right. \label{rubo} \\
  & & \left. - i \, {\rm sgn}(km) \sqrt{1-f^2} \left(
\begin{array}{cc}\sin(k\theta) \\
- \cos(k\theta) \end{array}
  \right) \right ]. \nonumber
\end{eqnarray}
Here, $m,k \in \{-1/2,+1/2\}$ select the rotation directions for the phase and polarization, respectively;  $n(\theta)$ is the effective one-dimensional density of the condensate, and $f$ is a real constant which gives the degree of circular polarization, and $|f|$ must be less than  $1$. In the Rubo vortex state, $f$ can depend on the radius $r$, while in a ring, we can approximate that $f$ is nearly constant.  
For each combination of $k$ and $m$, this ansatz gives a degree of circular polarization $c={\rm sgn}(km) 2 f \sqrt{1-f^2} $ which does not depend on $\theta$, and a linear polarization angle that rotates as $k\theta$. In the absence of interactions, and in a homogeneous ring $n(\theta)=n$, these states with $|f|=1/\sqrt{2}$ are eigenstates of the Hamiltonian that consists of the kinetic energy, 
\begin{equation}
H_{\rm kin}=-\frac{\hbar^2}{2m} \nabla^2 \equiv -\frac{\hbar^2}{2mR^2}\frac{d^2}{d\theta^2},
\end{equation}
where $m$ is the effective mass of the polaritons.

In a simply connected geometry, only half quanta of the type (\ref{rubo}) are allowed, because there is an implicit boundary condition that the polarization must be continuous at $r=0$. However, in a ring geometry, this condition is relaxed, and other wave functions with half-quanta of circulation also satisfy the boundary condition that the wave function be single-valued. The polarization pattern we observe is reproduced by using the following form for the wave function, which is the same except for a single sign change:
\begin{eqnarray}
\vec\psi_{k,m} (\theta)  =& & \sqrt{n(\theta)}e^{im\theta}\left [f \left(
\begin{array}{cc}\cos(k\theta) \\
 \sin(k\theta) \end{array}
  \right) \right.  \label{expansatz} \\
  & & \left. + i \, {\rm sgn}(km) \sqrt{1-f^2} \left(
\begin{array}{cc}\sin(k\theta) \\
 \cos(k\theta) \end{array}
  \right) \right ].\nonumber
\end{eqnarray} 
This gives the polarization map shown in Fig.~5(b), for the choice $f = -0.3.$ 
Like the Rubo state, it provides a half-quantum of circulation with a phase rotation direction chosen by the sign of $m$, but it involves a flip of the circular polarization around the ring.  For these states, the degree of circular polarization is given by $c=-\ {\rm sgn}(km)  2f \sqrt{1-f^2} \cos(2 k \theta)$, so that the circular polarization direction, i.e., the $z$-component of the spin-1 particles, flips from one side of the ring to the other, as seen in the experimental data of Figure~5(a). 
The linear polarization angle is given by
\begin{equation}
\frac{1}{2}\arg \left(\frac{f+{\rm sgn} (km) \sqrt{1-f^2}  \,e^{2 i k \theta}}{f \,e^{2 i k
   \theta}-{\rm sgn} (km) \sqrt{1-f^2} }\right),
\end{equation}
which can rotate in either direction, and this direction is determined by the sign of $k$ in the ansatz. Note that any sign of $k$ and $m$ can be paired here.  As discussed in the online supplementary information, this ansatz also reproduces the interference patterns which we observe. 

The state state seen here is not an eigenstate of kinetic energy. As discussed further in the supplementary material, the ansatz (\ref{expansatz}) which fits our experimental data corresponds to the superposition
\begin{eqnarray}
\vec\psi_{k,m}  = -\sqrt{2}( 0.95 \vec{\varphi}^0_{\frac{1}{2},\frac{1}{2}}+0.3 \vec{\varphi}^0_{\frac{1}{2},-\frac{1}{2}}),
\end{eqnarray}
where $\varphi^0_{k,m}$ are the Rubo states (\ref{rubo}) with $f=0$, which are chosen because they give us an orthogonal basis of states with well-defined physical properties. In other words, the observed state is a superposition of different half-quantized circulating states.  Although a superposition of different angular-momentum states may seem odd, there is no in-principle reason for the system to favor a pure angular momentum state over the observed spin-flipping circulating state. Both states satisfy the boundary conditions. The state  (\ref{expansatz}) is not allowed in a simply-connected geometry, however, because the kinetic energy would diverge at $r=0$. 

\begin{figure}[t]
\begin{center}
\includegraphics[width=0.5\textwidth]{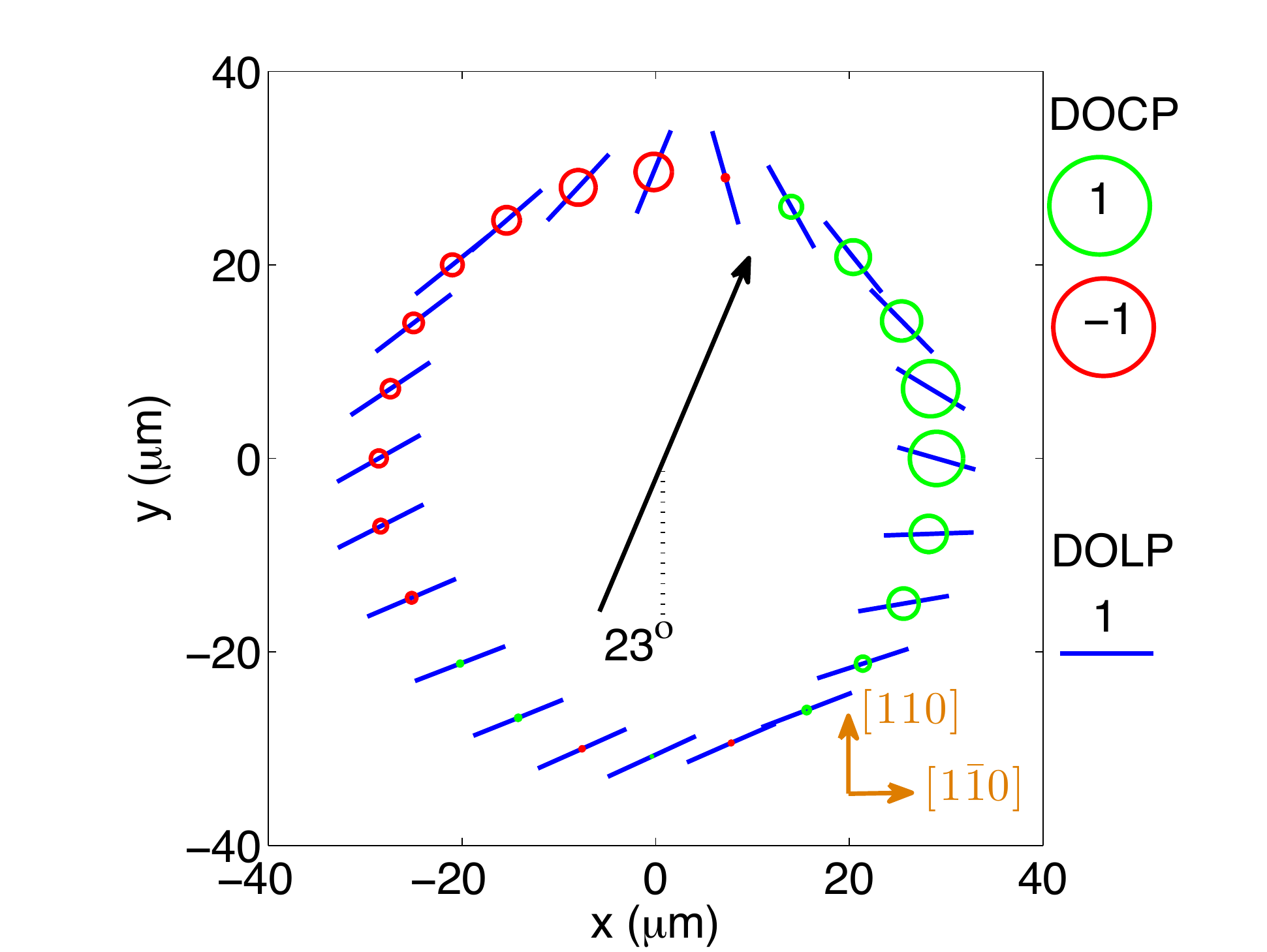}

\includegraphics[width=0.35\textwidth]{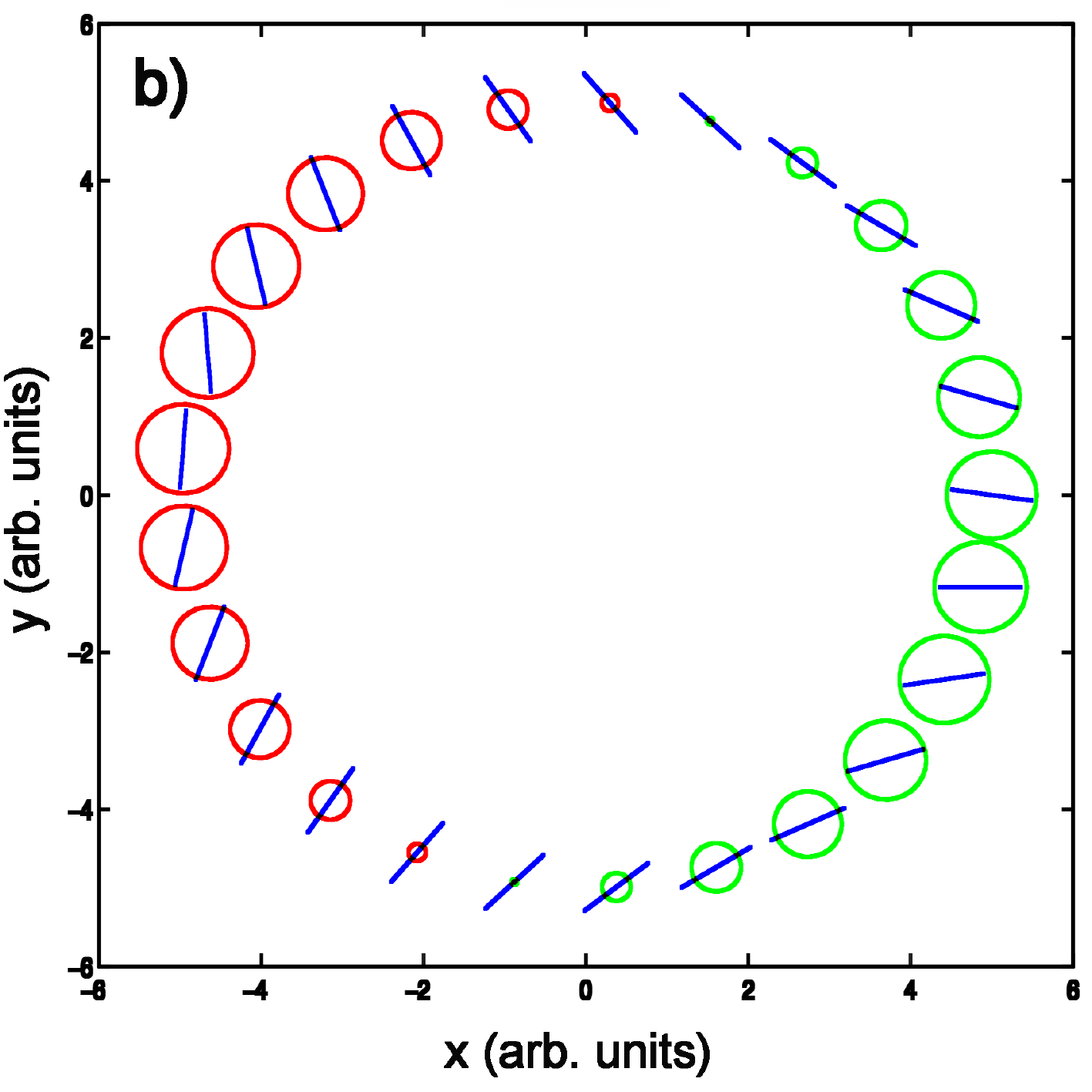}
 \end{center}
 \caption{a) Experimental results for the polarization of the light emission from the ring at various positions. The length of the solid blue lines is proportional to the degree of linear polarization, and the radius of the circles is proportional to the degree of circular polarization (green: right-handed; red: left-handed).  The arrow shows the direction of downhill flow of the polaritons in the gradient of the cavity. b) Theoretical polarization pattern for the condensate wave function given by (\ref{expansatz}), with $k=m=-1/2$ and $f=-0.3$. }
\label{pol}
\end{figure}

The fact that the stable state seen here is not an energy eigenstate is likely related to both the effect of particle-particle interactions as well as the dynamic quasi-steady-state conditions of generating the ring. We discuss the modifications which come from taking into account polariton-polariton interactions in the supplementary material.

\section{Discussion and Future Directions}

The stability of the polarization pattern observed in these experiments appears to be due to symmetry breaking related to the optical spin-Hall effect for polaritons \cite{kavOSH}. This effect comes about due to the energy difference of the transverse-electric (TE) and transverse-magnetic (TM) modes of the photons in the cavity at finite in-plane momentum. Polaritons moving under the force of the cavity gradient will preferentially have linear polarization orthogonal to their direction of motion. When these polaritons scatter elastically into other $k$-states with the same energy, they will scatter into a superposition of polarizations in the new direction, which will precess, giving a circular polarization component. As shown in Ref.~\cite{kavOSH}, the handedness of this circular polarization will be opposite for opposite directions of scattering. While this may be a small effect, it gives a symmetry-breaking term that corresponds to the circular polarization components being different on opposite sides of the downhill gradient, as we observe in these experiments. Below the condensate threshold, there is no significant circular polarization component of the photoluminescence from the ring. Above the condensation threshold, the high occupation of the condensate amplifies scattering processes, so that any small symmetry breaking term can lead to the condensate adopting a new symmetry.

Since the condensate must satisfy the boundary conditions of the ring, it cannot have the underlying four-fold rotational symmetry of the GaAs crystal unless it is in a much higher angular momentum state. The spin-flipping state with half-quantum of circulation which we see is actually well-matched to the real-space separation of the different circular handedness favored by the optical spin-Hall effect in the presence of the cavity gradient. The polarization rotation is therefore not a spontaneously broken symmetry. But because the same polarization pattern can satisfy the boundary conditions of the ring with either handedness $m = \pm \frac{1}{2}$ of the flow (phase gradient) around the ring, we see that the direction of the circulation does change randomly, as this degree of freedom has spontaneously broken symmetry. Future theoretical work will address the dynamical considerations of how the ring condensate forms and how it gains a circulation direction.  Under some circumstances, a non-circulating state can be unstable when generation and decay are accounted for, as, for example, the vortex which appears in a bathtub drain \cite{keeling}. 

Recent work with a laser-generated ring trap \cite{baum-ring} showed a superposition, i.e., a standing wave, or two counter-rotating polariton waves in high-momentum states. The main differences between that experiment and the work presented here is that the lifetime of the polaritons in our work is more about an order of magnitude larger than in the work of Ref.~\cite{baum-ring}. This allows the polaritons in the present work to cool down and thermalize to the bottom of the trap, as seen in Figure~3, which allows the condensate to be in the true ground state of the ring. In the work of Ref.~\cite{baum-ring}, the polaritons maintained the energy they had where they were generated by the laser. The momentum of the polaritons seen in Ref.~\cite{baum-ring} was therefore fixed by that initial energy. There was no evidence of spontaneous symmetry breaking of the circulation direction under those conditions.

Further experimental work in our ring traps can use a resonant laser beam to inject angular momentum unto the condensate; that is, to stir it. It is not clear whether injecting new particles with finite momentum will simply raise the amplitude of the condensate in its existing half-quantum state, or if the condensate will prefer to jump to a higher angular momentum state. Both are ways in which the condensate can increase its total momentum.  We also have the possibility of introducing small barriers in the ring using a laser-generated exciton cloud, to create Josephson junctions analogous to those used in a ring SQUID. Since we can observe the interference patterns directly from the light leaking through the mirrors, we can nondestructively measure the phase map for all the states we produce. 

As we have seen, the polarization rotation is pinned in this system while the direction of circulation of current around the ring is not. The laser focused at the center of the ring does not introduce any circulation. As a result, we see a random occurrence of circulation to the left or right, even as the polarization is pinned. These experimental results can therefore be seen as an example of spontaneous time-reversal symmetry breaking leading to a persistent current around the ring, with a phase coherence time at least a hundred thousand times longer than the lifetime of any one particle in the condensate. 

Because we have a macroscopic ring geometry which is topologically distinct from a simply-connected geometry, a different topology of the circulation is allowed, with the spin of the particles flipping around the ring, even as the particles remain in a single, macroscopic wave function. The spin-flipping state with quantized flow cannot be continuously transformed into a state with a full quantum of circulation or into a state with a half-quantum of circulation and no spin flip. The quantized angular momentum state seen here is different from the typical case of pairs of vortices of opposite vorticity generated due to turbulence \cite{dev-anti,marz-anti}, and also can be produced on demand, as opposed to needing to search for a pinned vortex at a random location in a disordered landscape, as was the case for Ref.~\cite{dev-half}.

\section*{Appendix. Supplementary Information}

{\bf Calculation of the  interference pattern}. As discussed in the main text, the wave function ansatz which gives the polarization pattern of the data has the azimuthal angle dependence
\begin{eqnarray}
\psi_{k,m} (\theta)  &= & \sqrt{n(\theta)}e^{im\theta}\left [f \left(
\begin{array}{cc}\cos(k\theta) \\
 \sin(k\theta) \end{array}
  \right) \right.  \label{expansatz} 
   \left. + i \, {\rm sgn}(km) \sqrt{1-f^2} \left(
\begin{array}{cc}\sin(k\theta) \\
 \cos(k\theta) \end{array}
  \right) \right ],
\end{eqnarray}
with $k=1/2$, $m = \pm 1/2$.  Here we show that this wave function gives the interference pattern we see. 

For simplicity we assume that the density $n(\vec{r})$ is independent of $\theta$ but has a Gaussian radial dependence, $n(r) = e^{-(r-r_0)^2/\sigma^2}$. 
For $k=m=1/2$, the full wave function can then be written as
\begin{eqnarray}
\psi(r,\theta) = \sqrt{n(r)}e^{i\theta/2}\left[ 
f\left(\begin{array}{c}\cos\theta'/2 \\ \sin\theta'/2\end{array}\right) 
+ i\sqrt{1-f^2}\left(\begin{array}{c}\sin\theta'/2 \\ \cos\theta'/2\end{array}\right)
  \right], \nonumber\\
  \label{full}
\end{eqnarray}
where $\theta' = \theta - \theta_0$, and $\theta_0$ is an offset angle. In the case shown in Fig.~5 of the main text, $\theta_0 = 23^\circ$. If $\theta_0 = \pi/2$, there will be no interference fringes because the two images will have crossed polarization at every point.
\begin{figure}
\includegraphics[width=0.7\textwidth]{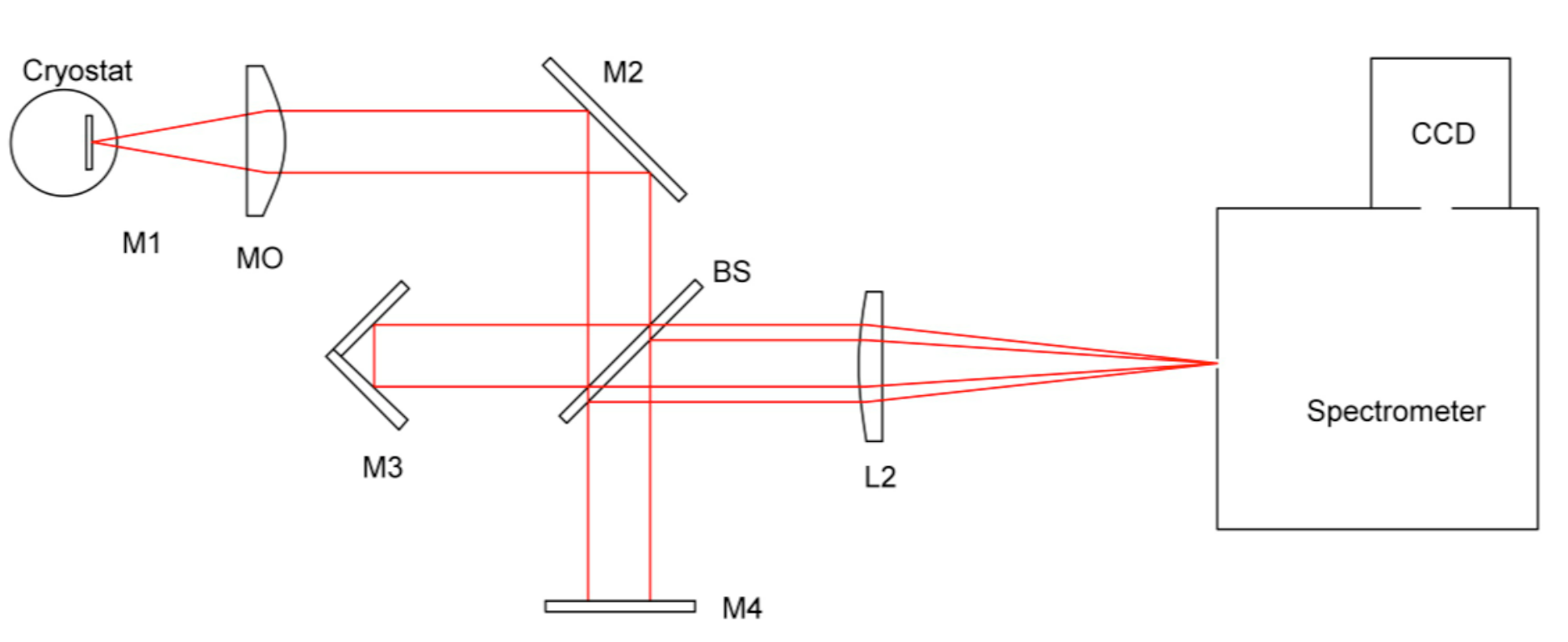}
\caption{Geometry of the interference measurement. M0: microscope objective. M2, M4: flat mirrors. M3: retroreflector. BS: beamsplitter. L2: imaging lens. The spectrometer is set to zero-order mode to spectrally integrate all of the light from the ring.}
\label{Sint}
\end{figure}

\begin{figure}
\includegraphics[width=0.4\textwidth]{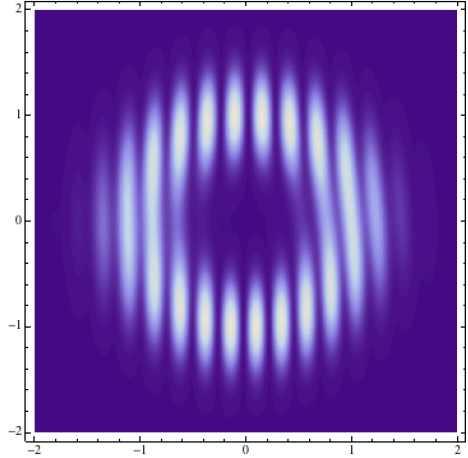}
\caption{Interference pattern for the wave function (3) of the main text, for an arbitrary ring width and constant density $n(\theta)$ around the ring, with $f = 0.3$ and $\theta_0 = 23^{\circ}$.}
\label{bigint}
\end{figure}

The reflected copy of this wave function will have $\theta \rightarrow \pi -\theta$, but will have the circular handedness reversed, since one copy of the image is reflected from one mirror in the Michelson interferometer, and the other is reflected two mirrors in a two-mirror retroreflector (see Fig.~\ref{Sint}), and each mirror flips the handedness of circularly polarized light. The wave function of the reflected pattern is therefore
\begin{eqnarray}
\psi_r(r,\theta) = -\sqrt{n(r)}e^{-i\theta/2}\left[ 
f\left(\begin{array}{c}-\cos\theta'/2 \\ \sin\theta'/2\end{array}\right) 
- i\sqrt{1-f^2}\left(\begin{array}{c}\sin\theta'/2 \\ -\cos\theta'/2\end{array}\right)
  \right], \nonumber\\
  \label{fullref}
\end{eqnarray}
To generate the interference pattern, one copy is assumed to enter the image plane with $+k_x$ in the $x$-direction, and the other with $-k_x$, so that the interference pattern is
\begin{eqnarray}
I(x,y) = | \vec\psi(x,y)e^{ik_x x} + \vec\psi(-x,y)e^{-ik_x x} |^2 
\end{eqnarray}
Fig.~\ref{bigint} shows the pattern generated using $r_0 = 1$, $\sigma = 0.5$, and $k_x = 2\pi/\lambda$ with $\lambda = 0.5$.  There is one more fringe on the bottom than on the top.  In this same reflection geometry, a state with $m = 1$ would have two more fringes on one side than on the other side.

\begin{figure}
\includegraphics[width=0.5\textwidth]{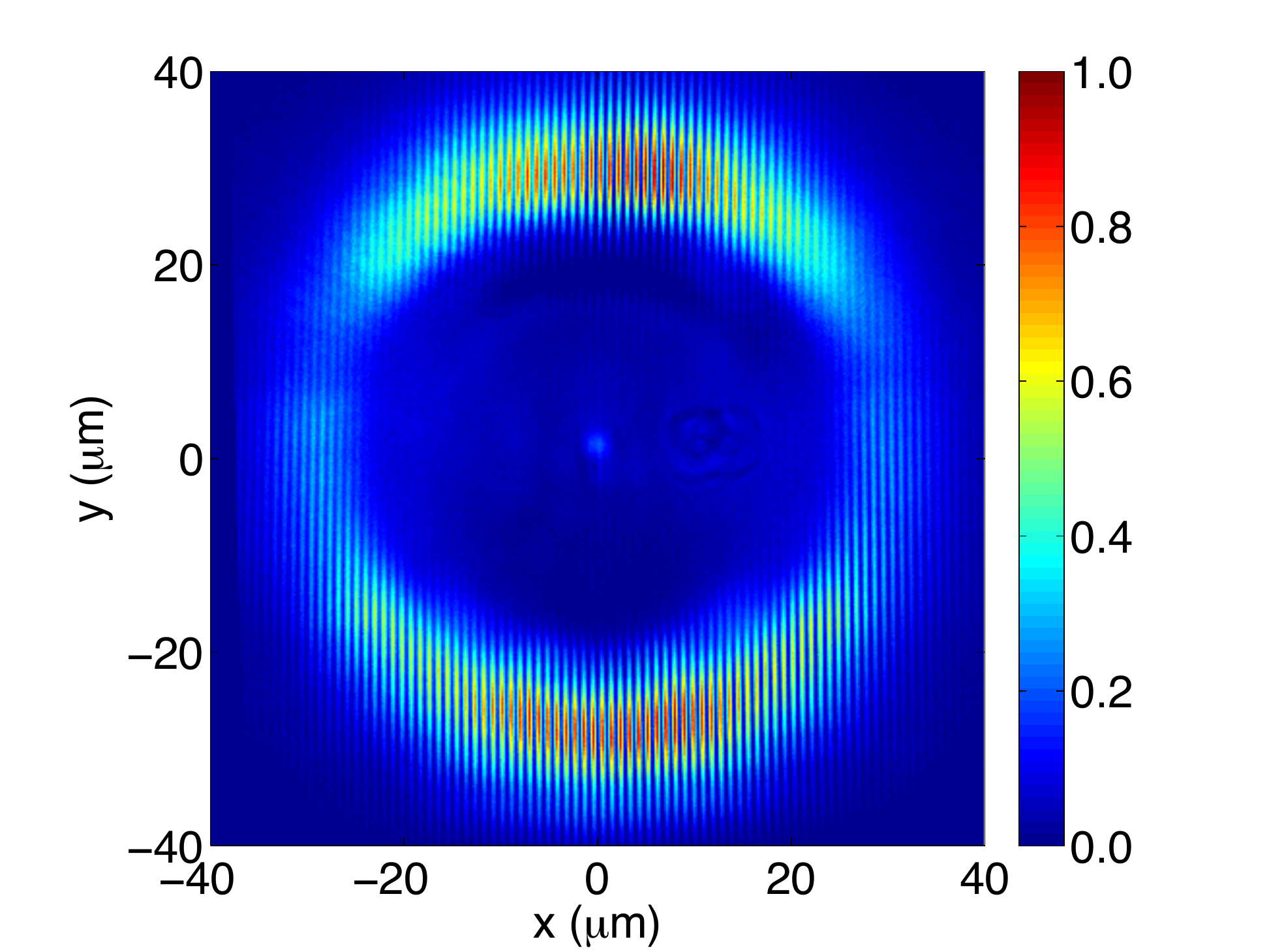}
\caption{Experimental interference pattern for the same conditions as the experiments of Figures 2(b) and 3(b) of the main text. In this case, there is one more fringe on the bottom than on the top.}
\label{newint}
\end{figure}

Figure~\ref{newint} shows a typical experimental interference pattern corresponding to the data of Figs.~2(b) and 3(b) of the main text.

{\bf Method of obtaining the phase map.} 
The interference pattern is first transformed into a reciprocal space map through by the 2D fast Fourier transform (FFT) method. In the reciprocal space map, there are two high-frequency components which correspond to the interfering part of the interference pattern, and a low-frequency component centered at $k=0$ which corresponds to the constant background of the interference pattern. The high-frequency components are complex conjugates of each other, carrying the phase information of the interference pattern. One of the high-frequency component sis chosen and shifted to the center of the reciprocal space map. By this, the frequency component that comes from the projection of the wave vector of the interfering beam on the image plane is removed. Then the 2D inverse FFT is applied to the filtered and shifted reciprocal map, from which the phase of the original ring condensate is obtained. Notice that in our set up, the phase difference between the ring condensate and its inverted image is detected. 

{\bf Additional details on the polarization state}. 
In order to determine the circular polarization of a state $\vec \psi = (\psi_1,\psi_2)$, we take the circularly polarized components $\psi_L=(\psi_1 + i\psi_2)/\sqrt{2}$ and $\psi_R=(\psi_1 - i\psi_2)/\sqrt{2}$. We then define the linear polarization angle as \[\frac{1}{2}\arg\left(\frac{\psi_R}{\psi_L}\right),\] and the degree of circular polarization 
\[c=\frac{|\psi_R|^2-|\psi_L|^2}{|\psi_R|^2+|\psi_L|^2}.\]

\begin{figure}
\includegraphics[width=0.4\textwidth]{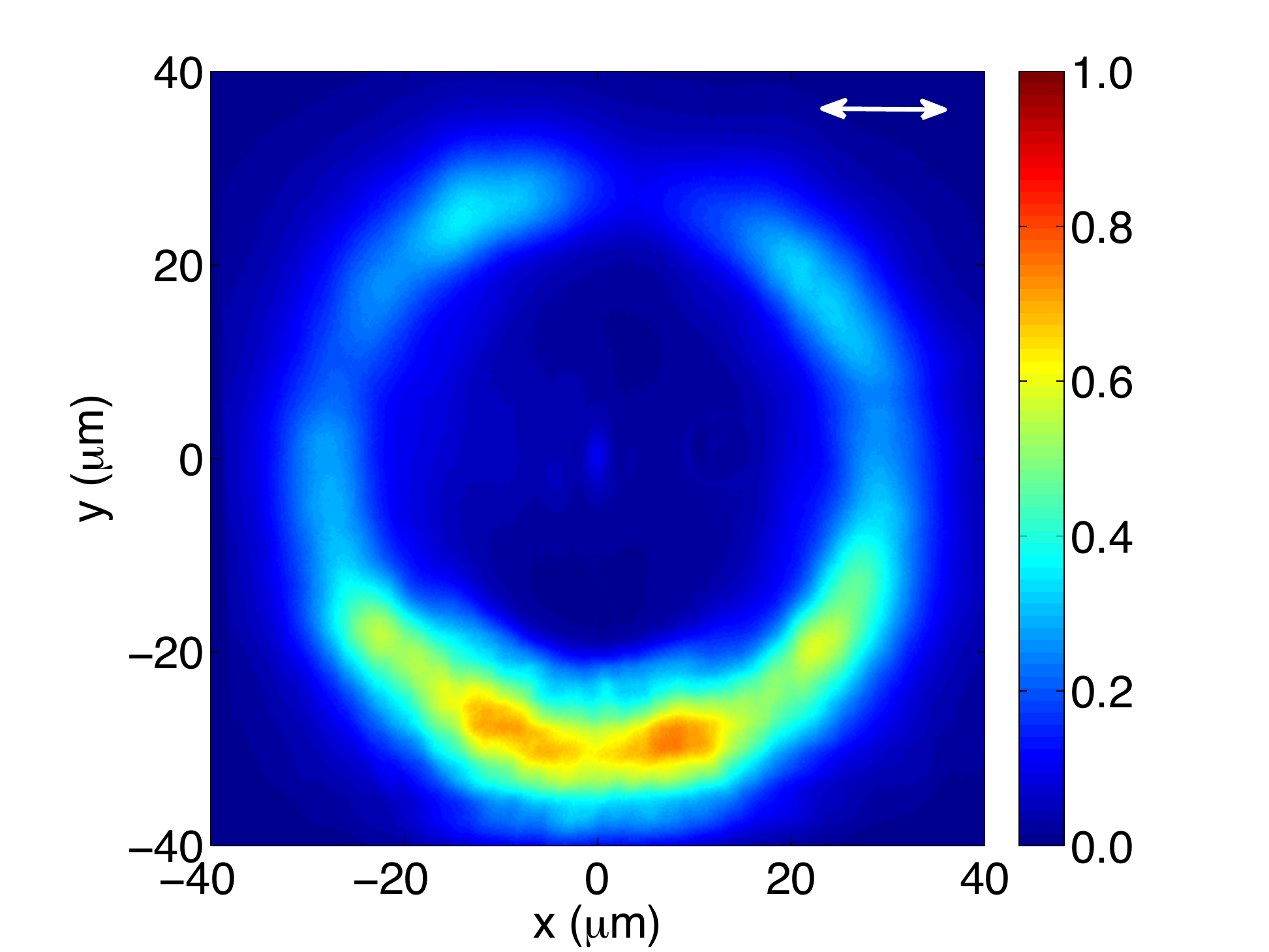}
\includegraphics[width=0.4\textwidth]{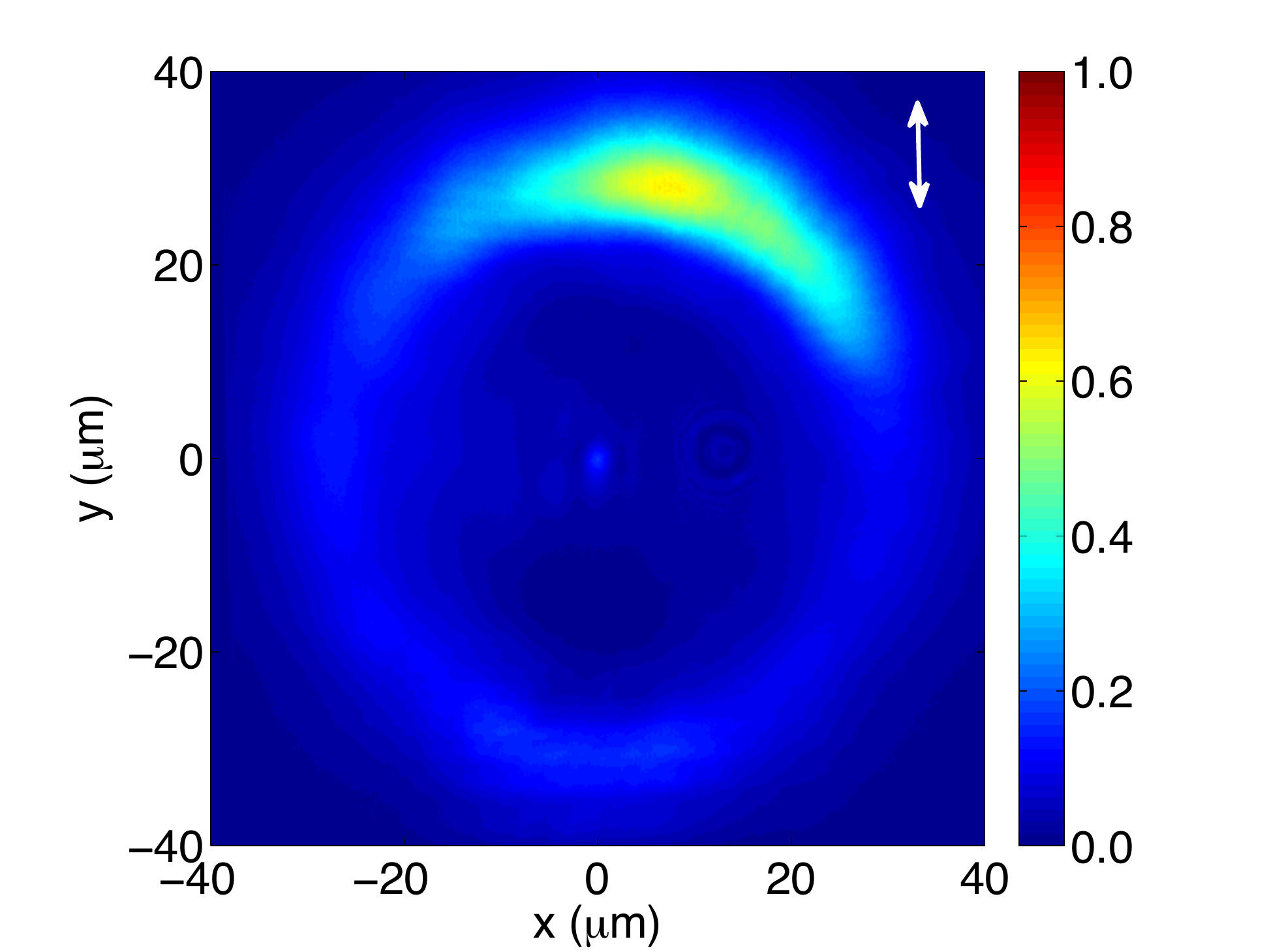}
\includegraphics[width=0.4\textwidth]{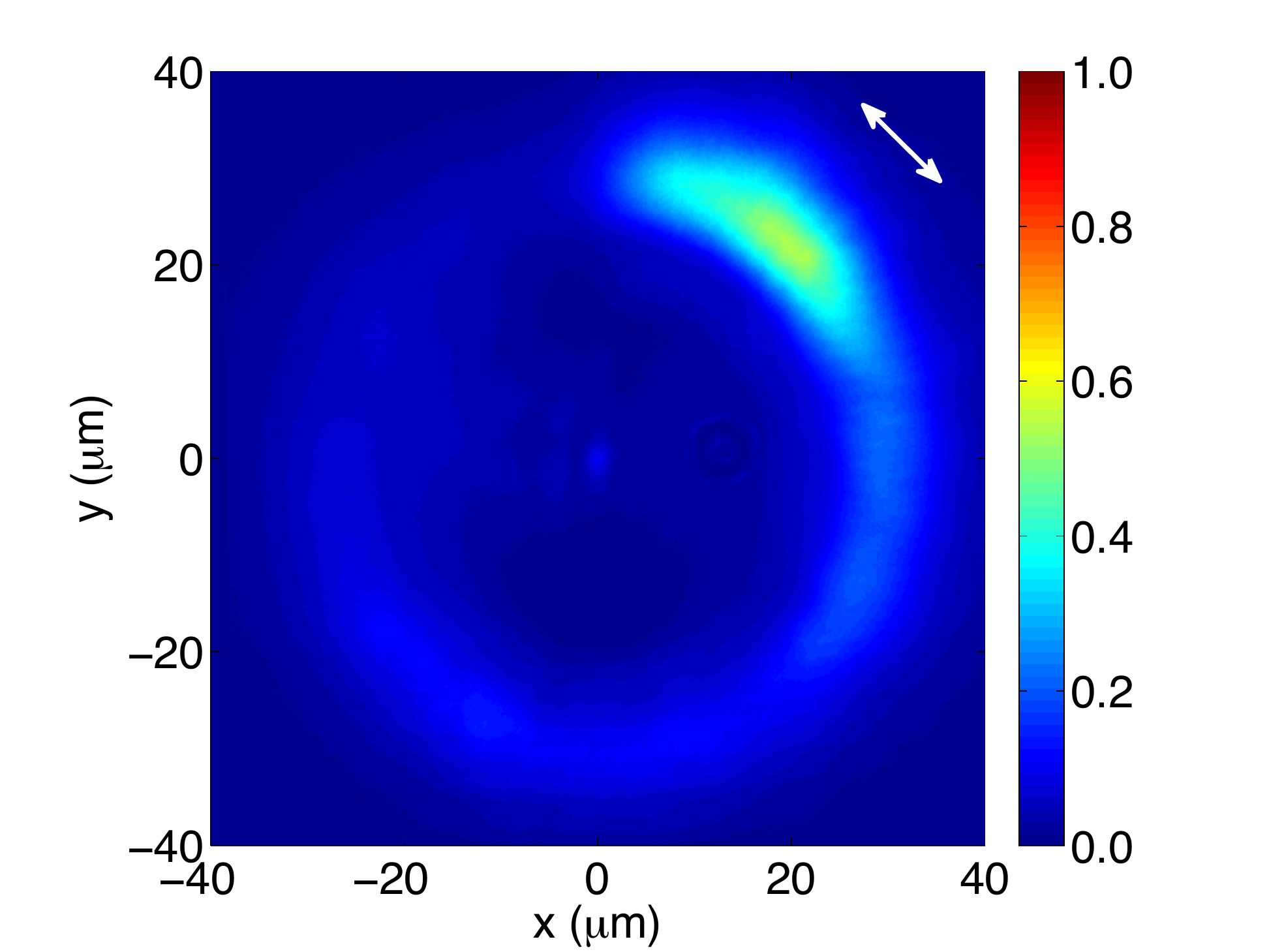}
\includegraphics[width=0.4\textwidth]{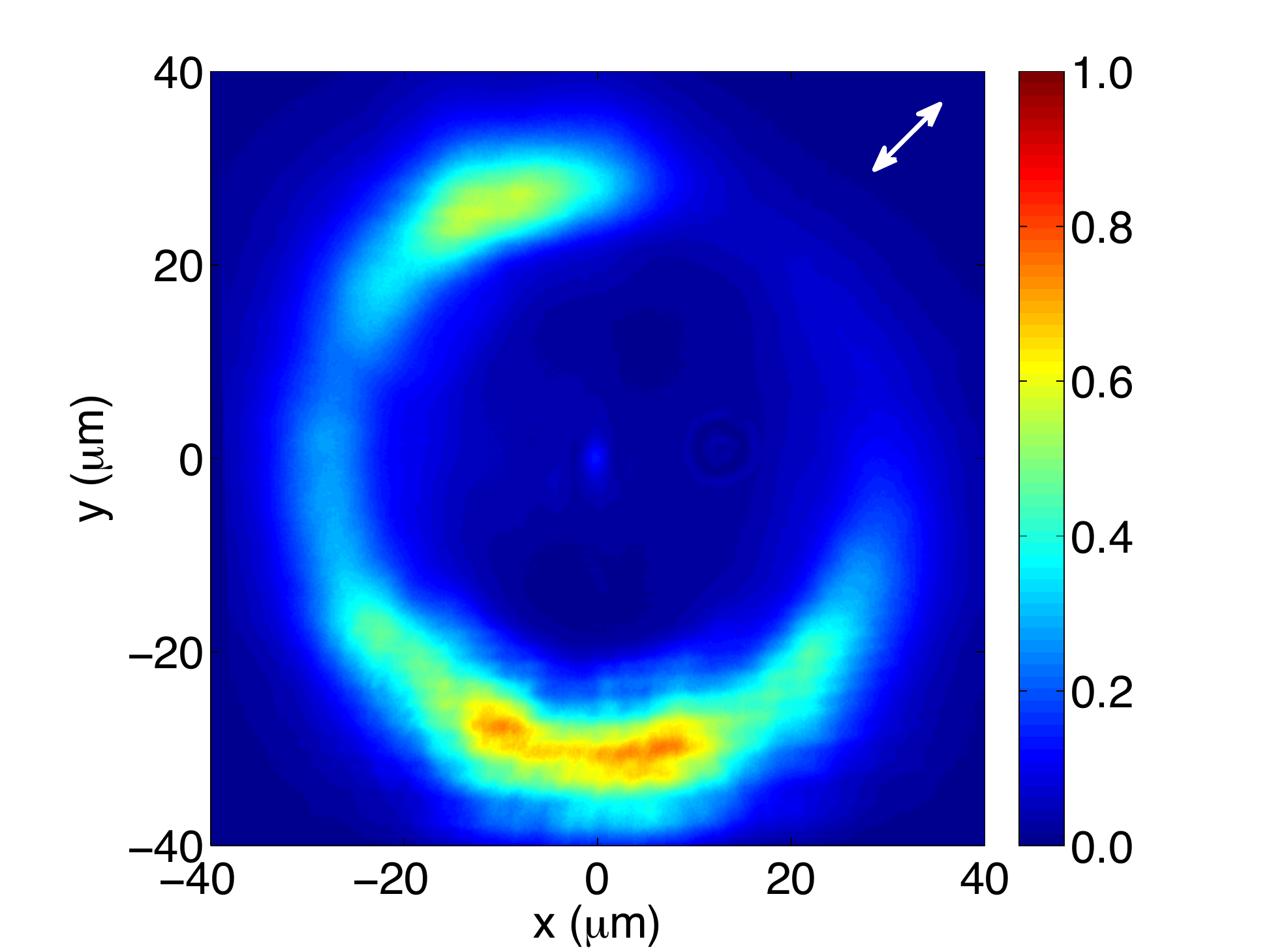}
\includegraphics[width=0.4\textwidth]{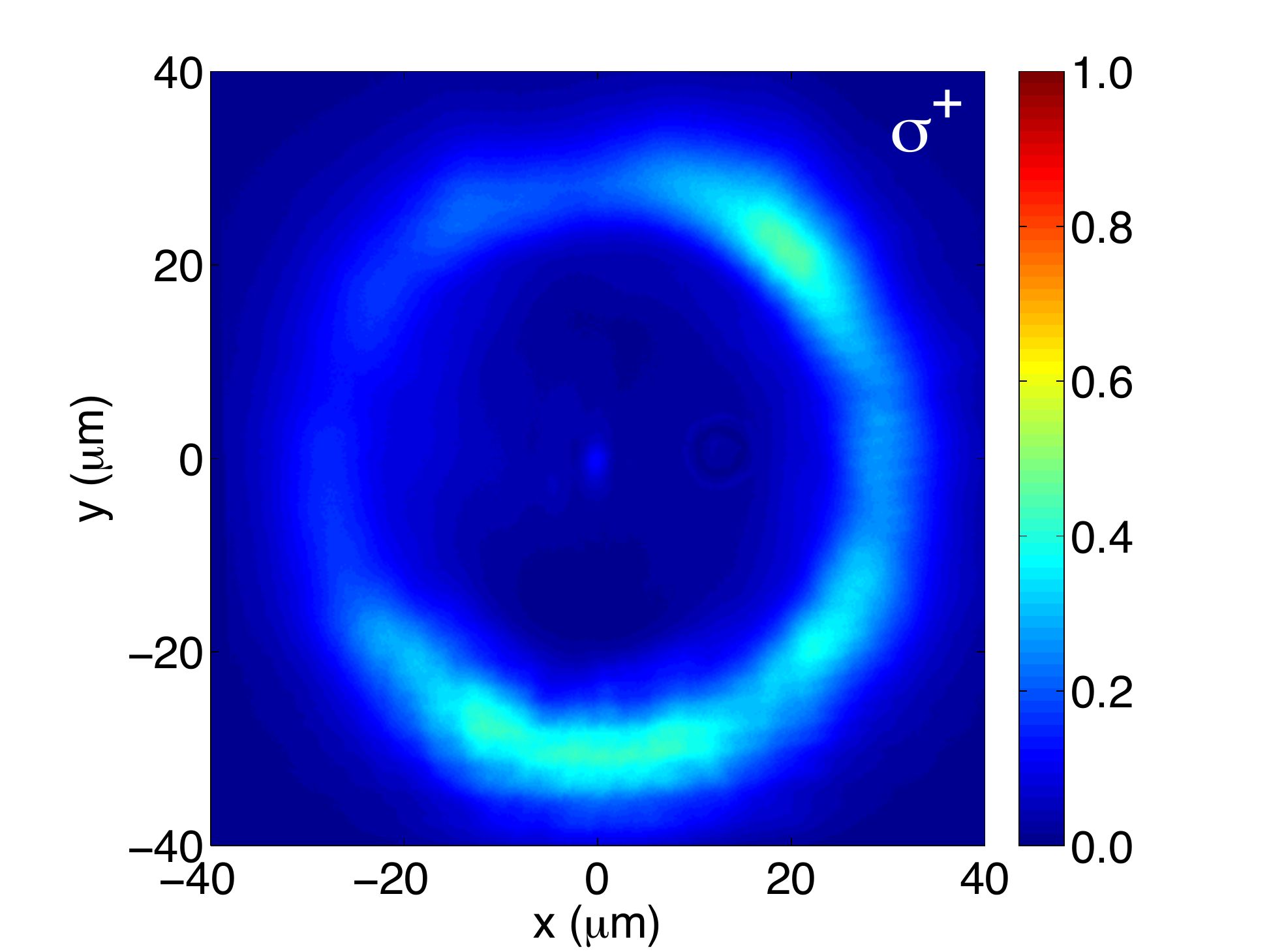}
\includegraphics[width=0.4\textwidth]{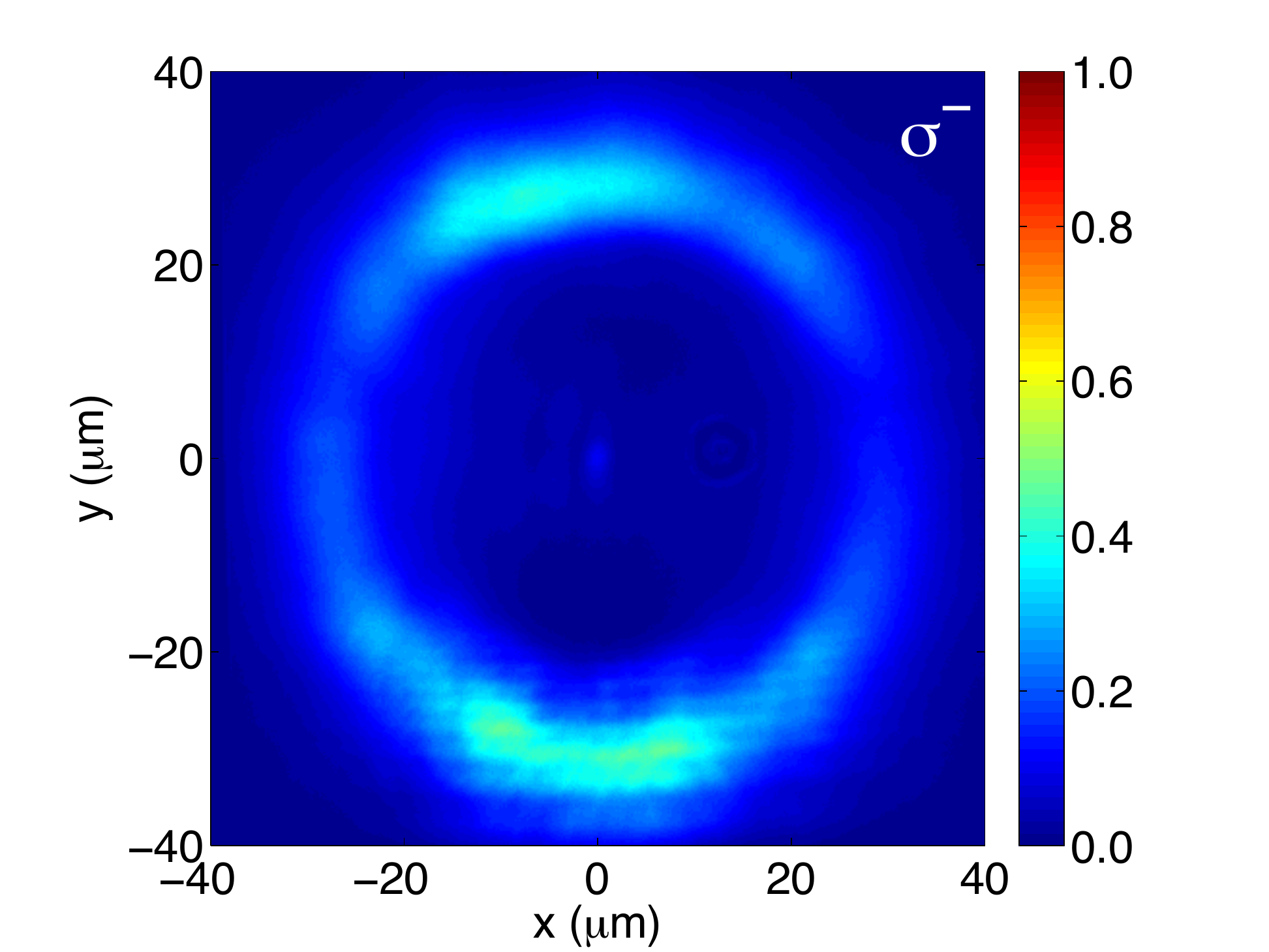}
\caption{Polarization-resolved images of the ring used for the data of Fig.~5(a) of the main text.}
\label{stokes}
\end{figure}

The experimental polarization state is determined by measuring the full Stokes vectors for the image of the ring. Figure~\ref{stokes} shows typical data, corresponding to the data of Figure~5(a) of the main text.

The Rubo half vortex (Equation (1) of the main text) is an eigenstate of the kinetic energy, but is not the only state with half-quantum vorticity which satisfies the boundary conditions. The spin-flipping half-quantum state which reproduces the experimental pattern (Equation (3) of the main text) can be seen as a superposition of elementary Rubo states. If we set $f=0$ we get four linearly independent half-vortex states for the Rubo states, which can be viewed as ``pure'' Rubo states:
\begin{equation}
\vec{\varphi}^{ 0}_{k,m}= {\rm sgn}(km)\frac{i}{\sqrt{2}}e^{{\rm sgn}(m) i\theta/2}\left(\begin{array}{c}- \sin\theta/2\\ {\rm sgn}(k) \cos\theta/2 \end{array}\right).
\end{equation}

The state shown in Figure 5 of the main text corresponds to
\begin{eqnarray}
\vec{U} &=& 0.95ie^{i\theta/2} \left(\begin{array}{c} \sin\theta/2\\ \cos\theta/2 \end{array}\right) -0.3e^{i\theta/2}\left(\begin{array}{c} \cos\theta/2\\ \sin\theta/2 \end{array}\right) \nonumber \\
&=& 0.95ie^{i\theta/2} \left(\begin{array}{c} \sin\theta/2\\ \cos\theta/2 \end{array}\right) 
-0.3ie^{i\theta'/2}\left(\begin{array}{c} -\sin\theta'/2\\ \cos\theta'/2 \end{array}\right) \nonumber \\
&=&-\sqrt{2}( 0.95 \vec{\varphi}^0_{\frac{1}{2},\frac{1}{2}}+0.3 \vec{\varphi}^0_{\frac{1}{2},-\frac{1}{2}}),
\end{eqnarray}
where $\theta' = \theta + \pi$.

{\bf Effect of interactions}.
The energy of the states can be evaluated using a variational approach in an effective 1D model, determining $f$ by minimizing the total energy $E=E_{\rm kin}+E_{\rm int}$, consisting of the kinetic energy 
\begin{equation}
E_{\rm kin}=\frac{1}{2\pi} \int_0^{2 \pi}d\theta \ \vec\psi^* (\theta) \cdot \left[-\frac{\hbar^2}{2mR^2}\frac{d^2}{d\theta^2} \right] \vec\psi(\theta)
\end{equation}
and the interaction energy \cite{rubo}
\begin{equation}
E_{\rm int}=\frac{1}{4\pi} \int_0^{2 \pi}d\theta \ \left[(U_0-U_1)(\vec\psi^* \cdot \vec\psi)^2 + U_1|\vec\psi^* \times \vec\psi|^2\right].
\end{equation}
Here, $U_0$ and $U_1$ are the interaction constants, which can be written in terms of the interactions between the same circular polarization $M_{\uparrow \uparrow}$ and different circular polarizations $ M_{\uparrow \downarrow}$, as $U_0=M_{\uparrow \uparrow}$, $U_1$=$M_{\uparrow \uparrow}-M_{\uparrow \downarrow}$. Relative to $M_{\uparrow \uparrow}$, $ M_{\uparrow \downarrow}$ is small and is attractive rather than repulsive.  

Strong interactions favor linear polarization at high density.  Therefore when the density around the ring is varied, e.g. by moving the center spot slightly, the polarization pattern shifts. However, the overall polarization pattern is stable, with the separation of left- and right-circular components relative to the gradient of the cavity always the same as shown in Figure~5(a) of the main text. The degree of circular polarization is always stronger on the downhill side of the gradient. This is consistent with the picture that polaritons moving uphill will take longer to reach the condensate, and therefore have more spin randomization.

\begin{figure}
\includegraphics[width=0.6\textwidth]{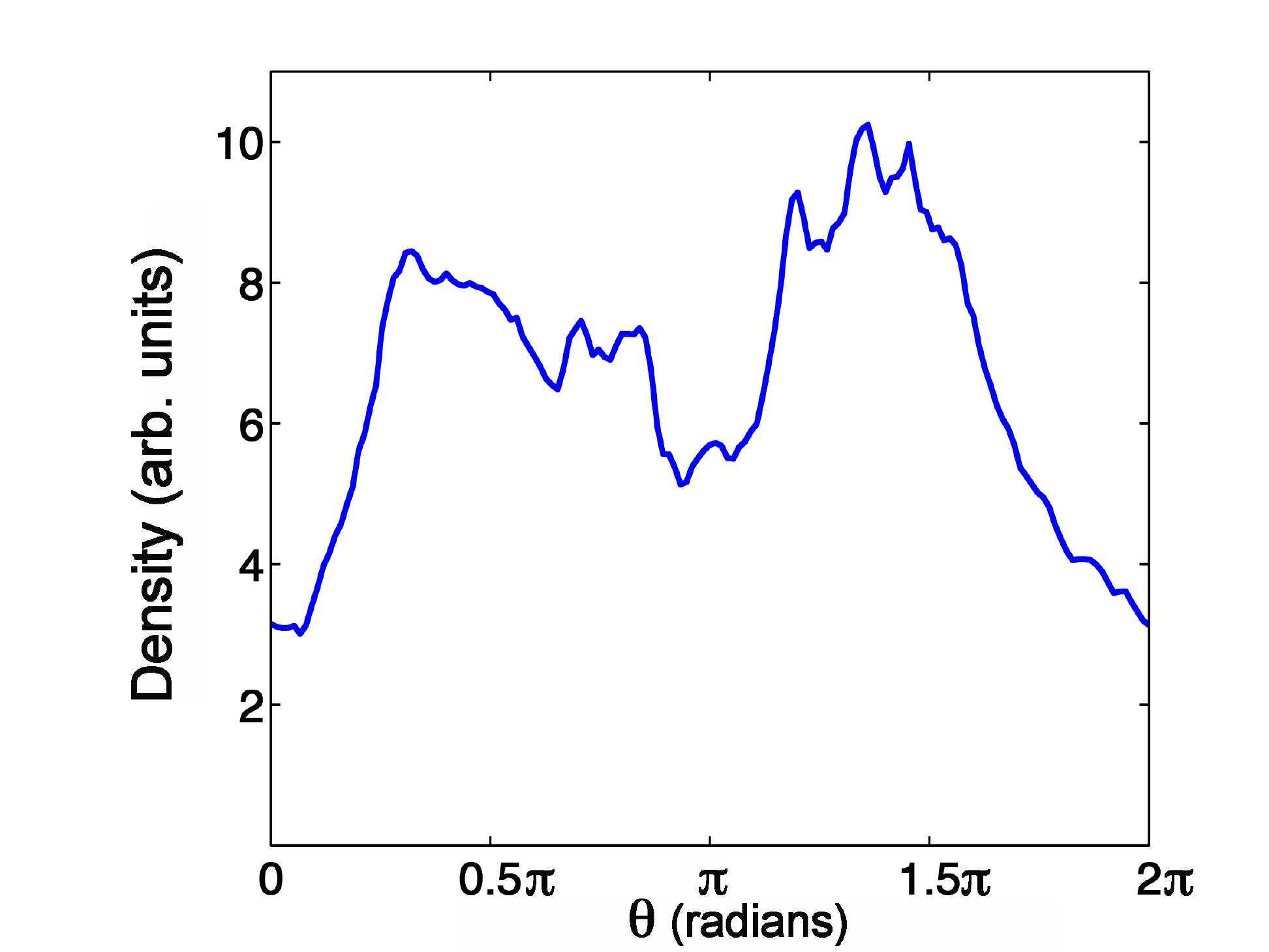}
\caption{Relative density as a function of angle around the ring, $n(\theta)$, for the same conditions as the data of Fig.~5(a) of the main text. The angle $\theta$ is measured relative to the [1$\bar{1}$0] axis, i.e. the $x$ axis on the plots of the ring.}
\label{dens}
\end{figure}

The relative density of the condensate around the ring can be measured by taking the total light emission, spectrally integrated and summing polarizations, as a function of angle. Figure~\ref{dens} shows a typical plot, corresponding to the conditions of Figure~5(a) of the main text. 

\begin{acknowledgments}
This work has been supported by the National Science Foundation under grants DMR-1104383 and PHY-1148957. The work at Princeton was partially funded by the Gordon and Betty Moore Foundation as well as the National Science Foundation MRSEC Program through the Princeton Center for Complex Materials (DMR-0819860). 

\end{acknowledgments}

\end{document}